# Scaling of the spin Seebeck effect in bulk and thin film

By K. Morrison,[1]* A.J Caruana,[1,2] C. Cox[1] & T.A. Rose[1]

[*]    Dr K. Morrison
[1]Department of Physics, Loughborough University,
Loughborough LE11 3TU (United Kingdom)
[2]ISIS Neutron and Muon Source, Didcot, Oxfordshire, OX11 0QX
E-mail: k.morrison@lboro.ac.uk

Whilst there have been several reports of the spin Seebeck effect to date, comparison of the absolute voltage(s) measured, in particular for thin films, is limited. In this letter we demonstrate normalization of the spin Seebeck effect for $Fe_3O_4$:Pt thin film and YIG:Pt bulk samples with respect to the heat flux $J_q$, and temperature difference $\Delta T$. We demonstrate that the standard normalization procedures for these measurements do not account for an unexpected scaling of the measured voltage with area that is observed in both bulk and thin film. Finally, we present an alternative spin Seebeck coefficient for substrate and sample geometry independent characterization of the spin Seebeck effect.

## I. INTRODUCTION

The spin Seebeck effect (SSE) is defined as the production of a spin polarized current in a magnetic material subject to a thermal gradient. It was first highlighted by Uchida *et al.* [1], and led to the development of a new field of magnetothermal effects (such as spin Peltier and spin Nernst) now grouped together under spin caloritronics [2].

Observation of the spin Seebeck effect is typically achieved by placing a heavy metal such as Pt in contact with the magnetic material, where the spin current generated in the magnetic layer is injected into the Pt layer and converted to a useable voltage by the inverse spin Hall effect [3,4]. The advantage of the SSE is that it can be used as a source of pure spin polarized current for spintronics applications, paving the way for a host of new devices (such as spin Seebeck based diodes) [5–8]. In addition to its potential applications in spintronics, it has been presented as an alternative to conventional thermoelectrics for harvesting of waste heat, due to decoupling of the electric and thermal conductivities that typically dominate the efficiency [9] [10] [11]. This is in part due to a completely different device architecture, such as that shown in Figure 1, where the magnetic layer can be chosen to have low thermal conductivity independent of the second paramagnetic layer, which in turn can be selected to have low resistivity.

In this work, we discuss the longitudinal SSE (LSSE) geometry (Figure 1(a)), since it lends itself readily to energy harvesting. An important factor to consider in this geometry is the impact of other magneto-thermal contributions to the measured voltage, $V_{ISHE}$, when assessing the LSSE. This includes the anomalous Nernst effect (ANE) and proximity induced ANE (PANE) in the Pt detection layer. For semiconducting or metallic thin films, such as presented here, both the ANE and PANE should be considered. For insulating magnetic materials, such as YIG, the PANE is the only potential artefact of the measurement.

Whilst there have been extensive studies of the spin Seebeck effect over the last 5-10 years, this has predominantly been in the YIG:Pt system, where for thin films, the substrate is often GGG [14] [15]. Reports of the spin Seebeck coefficient for this material system can, however, vary significantly [9]. Whilst this is sometimes attributed to the quality of the interface [14] there are indications that it could also be affected by an artefact of the measurement of $\Delta T$ [16], [17], [18].

Despite indications of the importance of heat flux in measurements [16], [17] [19], typically, LSSE measurements are still characterized by measuring the voltage generated across two contacts (of a spin convertor layer), as the temperature difference across the substrate, magnetic layer, and Pt is monitored. This is problematic, especially for thin films, where the temperature difference across the active material is not directly measured.

If such spin Seebeck devices are to be considered for power conversion there needs to be a shift in metrology so that meaningful comparison of the material parameters can be made. This is of particular importance if we consider the impact that the quality of the magnet:Pt interface can have on the observed voltage (i.e. efficiency of spin injection) [14]. An intermediate step is to use a simple thermal model to estimate the temperature difference across the active (magnetic) layer, however, this is limited by the unknown thermal conductivity of the thin films (likely to differ from the bulk), and thermal interface resistances. By measuring the voltage generated with respect to the heat flux, $J_Q$, we can obtain a robust measurement of the temperature gradient across the active material ($\nabla T = J_Q/\kappa$). In this case we are only limited by our knowledge of the thermal conductivity, $\kappa$.

In this paper we present measurement of the spin Seebeck voltage with respect to the temperature difference, $\Delta T$, and heat flux, $J_Q$. We demonstrate that the $J_Q$ method is indeed more reliable than the $\Delta T$ method and can be used to compare thin films with varying substrate thicknesses and thermal conductivity. In addition we find the surprising result that the measured spin Seebeck voltage, $V_{ISHE}$, is proportional not only to the contact separation, $L_y$, and temperature gradient, $\nabla T (= J_Q/\kappa)$, but also the sample area, $A_T$. This manifests as an increase in $V_{ISHE}/L_y\nabla T$ in both thin film and bulk samples with area, suggesting that the spin Seebeck coefficient typically used to define the magnitude of the effect in a bilayer system needs to be redefined. We present examples of this scaling in both thin film and bulk samples and put forward a new coefficient suitable for normalizing the measured voltage for generic sample geometries.

## II. EXPERIMENTAL METHOD

The thin film samples tested here were part of a series of 80:5 nm thick $Fe_3O_4$:Pt bilayers deposited using pulsed laser deposition (PLD) in ultra-high vacuum (base pressure 5 x $10^{-9}$ mbar) onto 0.15 mm, 0.3 mm and 0.5 mm thick borosilicate glass and 0.6 mm or 0.9 mm fused silica substrates. The laser fluence and substrate temperature for $Fe_3O_4$ and Pt deposition were 1.9 ± 0.1 J $cm^{-2}$ and 3.7 ± 0.2 J $cm^{-2}$, and 400 °C and 25 °C, respectively. $Fe_3O_4$ was chosen due to its large spin polarization (~80% [20]), low thermal conductivity (at 300 K: $\kappa_{thin\,film}$~2-3.5 W $m^{-1}$ $K^{-1}$, [21] $\kappa_{bulk}$ ~ 2-7 W $m^{-1}$ $K^{-1}$ [22]), and relative abundance of the constituent elements. Of particular note are the following: X-ray reflectivity data indicated a typical Pt thickness across the series of 4.5±0.5 nm and a roughness $\sigma$ = 1.5 ± 0.2 nm. $Fe_3O_4$ thickness was varied from 20 to 320 nm. In addition, these films have been shown to demonstrate highly textured, columnar growth, with grain sizes of the order of 100 μm. Further details are given in [19] and the supplementary information.

Bulk YIG was prepared by the solid state method [23]. Stoichiometric amounts of $Y_2O_3$ and $Fe_2O_3$ starting powders (Sigma Aldrich 99.999% and 99.995% trace metals basis, respectively) were ground and mixed together before calcining in air at 1050 °C for 24 hours. Approximately 0.5g of the calcined powder was then dry pressed into a 13 mm diameter, 1.8±0.2 mm thick cylindrical pellet. The pellet was then sintered at 1400 °C for 12 hours, after which, it was checked by XRD prior to sputtering 5 nm of Pt onto the as prepared surface using a benchtop Quorum turbo-pumped sputter coater. Samples were cut to size thereafter, using an IsoMet low speed precision cutter. Average grain size for this polycrystalline sample (determined using scanning electron microscopy) was 14.5 μm, with an open porosity of 30% measured by the Archimedes method. The thermal conductivity of this sample was measured using a Cryogenic Ltd Thermal Transport Option and found to be 3 W/K/m. Further details are given in the supplementary information. Magnetic characterization was obtained using a Quantum Design Magnetic Property Measurement System (SQuID) as a function of temperature and field. Magnetometry of the $Fe_3O_4$ films at room temperature typically exhibited a coercive field $H_c$ = 197±5 Oe, saturation magnetization $M_s$ = 90 emu $g^{-1}$, and a remanent moment of $M_r$ = 68 emu $g^{-1}$. Magnetometry of the YIG pellet exhibited a coercive field $H_c$ = 7 Oe, saturation magnetization $M_s$ = 25 emu $g^{-1}$, and a remanent moment of $M_r$ = 3 emu $g^{-1}$

Spin Seebeck measurements were obtained from a set-up similar to that of Sola et al. [16] and optimized for 12x12, and 40x40 mm samples. The thin film was sandwiched between 2 Peltier cells, where the top Peltier cell (1) acted as a heat source, and the bottom Peltier cell (2) monitored the heat $Q$, passing through the sample. Additional measurements were obtained in set-ups optimized for 10x10, 20x20 and 40x40 mm samples, where both the top and bottom Peltiers monitored the heat passing through the sample, and the heat source was a resistor mounted to the top side of the top Peltier. The Peltier cells were calibrated by monitoring the Peltier voltage $V_P$, generated as a current was passed through a resistor fixed to the top side of the Peltier cell (where the sensitivity $S_p$ of the set-ups varied from 0.11 to 0.264 V $W^{-1}$ at 300K). Two type E thermocouples mounted on the surface of the Peltier cells (such that they are in contact with the sample during the measurement) monitored the temperature difference across the sample, $\Delta T$; a schematic of this set-up is given in Figure 1(c). Further information on calibration is given in the supplementary information.

The voltage generated by the inverse spin Hall effect, $V_{ISHE}$, was determined by taking the saturation values at positive and negative fields, as shown in Figure 1(f) and demonstrated in detail in the supplementary information. This was repeated at several different heating powers, where $V_{ISHE}$ is expected to increase linearly with $\Delta T$ and $Q$. Finally, $\Delta T$ and $Q$ were plotted as a function of one another (see the example in Figure 1(e)), where non-linearity starts to appear if radiation losses become significant.

The impact of the ANE and PANE on the measured voltage was also taken into consideration. For the $Fe_3O_4$:Pt thin films we demonstrated that these contributions were negligible in a previous work by measuring thin films with a Au spacer layer between the $Fe_3O_4$ and the Pt. In this case, the $V_{ISHE}$ was similar to corresponding PM layer thicknesses without the spacer layer[19]. In addition, a separate work by Ramos *et al.*, has shown that the ANE contribution was ~3% of the total signal in their epitaxially grown $Fe_3O_4$ thin films and obtained an upper limit of contribution due to PANE of 7.5 nV/K.[12] This was an order of magnitude smaller than their observed $V_{ISHE}$, which is comparable to the measurements shown here. For the bulk YIG measurements, the ANE is considered negligible due to the insulating nature of the YIG and only the PANE in the Pt layer is a potential contribution to $V_{ISHE}$. In this case, it has been shown by several authors that the PANE in a >3 nm Pt layer on YIG is negligible with respect to the LSSE, by, for example, measurement in alternate geometries to separate the two contributions [13] or by introducing spacer layers between the YIG and the Pt. [24]

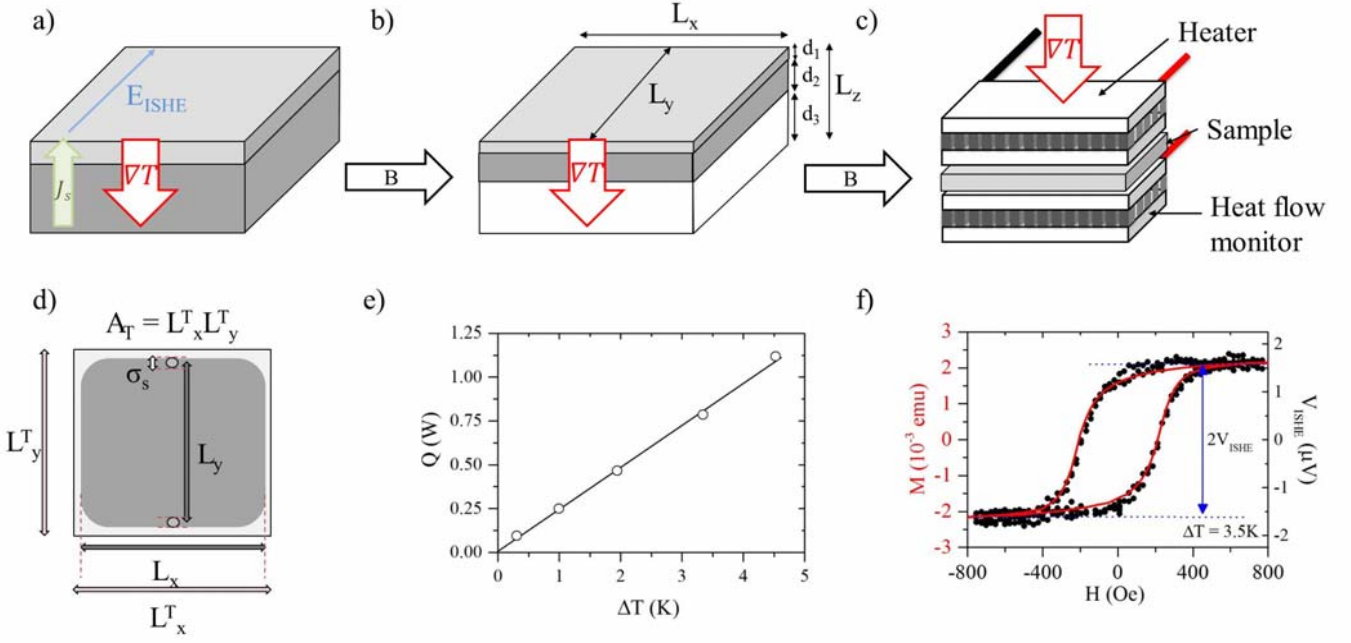

Figure 1. (a) Longitudinal spin Seebeck measurement geometries. Thermally generated spin current $J_s$, is produced in the magnetic layer (dark grey) and converted to a measureable voltage ($E_{ISHE}$) in the spin convertor layer (light grey) by the inverse spin Hall effect. (b) Typical longitudinal spin Seebeck thin film device; the length scales $L_x$, $L_y$, and $L_z$ denote device dimensions with respect to the thermal gradient (along z axis). Individual thicknesses of each layer $\{d_1, d_2, d_3\}$ are also indicated. (c) Schematic of the experimental set-up used in this work. (d) Top view of a typical thin film device where active material may not cover the entire substrate. In this case, the thermal contact area is quantified by lengthscales $L^T_x$ and $L^T_y$, active material width by $L_x$, and contact separation by $L_y \pm \sigma_s$, where $\sigma_s$ denotes contact size. (e) Example of the linear relationship between heat flux, $J_Q$, and temperature difference, $\Delta T$, in these measurements. (f) Example spin Seebeck data, $V_{ISHE}$ (symbols) from 80 nm $Fe_3O_4$:Pt thin film plotted alongside corresponding SQuID magnetometry (line).

### III. THEORY

It is useful at this point to note some key characteristics of the SSE measurement. First, the voltage has been shown (for a single device) to increase linearly with temperature difference ($\Delta T$) [10] or heat flux ($J_Q$) [16]. Secondly, it is known to increase linearly with contact separation ($L_y$) [10]. Lastly, for thin films, there have been indications of a length scale of the order of the magnon free path length above which the voltage generated saturates [25] [26].

In response to these observations, some of the first attempts to quantify the SSE were to normalize the voltage measured, $V_{ISHE}$, to the temperature difference $\Delta T$,

$$S_1 = \frac{V_{ISHE}}{\Delta T} \qquad (1)$$

where the units are (μV K$^{-1}$). More accurately, this could also be normalized to contact separation $L_y$,

$$S_2 = \frac{V_{ISHE}}{L_y \Delta T} \qquad (2)$$

with units of (μV K$^{-1}$ m$^{-1}$).

Given that the spin Seebeck effect is often defined in terms of the thermal gradient across the magnetic material $\nabla T$, the spin Seebeck coefficient is more often defined as,

$$S_3 = \frac{-E_{ISHE}}{\nabla T} = \frac{V_{ISHE} L_z}{L_y \Delta T} \qquad (3)$$

where the units are ($\mu$V K$^{-1}$), and the thermal gradient $\nabla T$ can be described by the temperature difference $\Delta T$, divided by the thickness of the sample $L_z$.

It was shown recently by Sola *et al.*, that there is the added complication of thermal resistance between the sample and the hot and cold baths (i.e. the interface across which $\Delta T$ is measured) [17]. Here they showed in measurements of the same sample in an experimental set-up at both INRIM and Bielefield that the measurement as a function of $\Delta T$ was unreliable – differing by a factor of 4.6 (gave $S_3$ = 0.231 $\mu$V K$^{-1}$ and 0.0496 $\mu$V K$^{-1}$, respectively), whereas by normalizing to heat flux, both set-ups obtained the same value to within 4% ($S_3$ = 0.685 and 0.662 $\mu$V K$^{-1}$, respectively).

In this work they measured the heat flux, $J_Q$, using a calibrated Peltier cell and initially determined the normalized voltage generated per unit of heat flux,

$$S_4 = \frac{V_{ISHE}}{L_y \left( \frac{Q}{A_T} \right)} \quad (4)$$

where $Q$ is the heat passing through a sample with cross-sectional area surface $A_T$, and the units are ($\mu$V m W$^{-1}$). Given that the thermal conductivity can be defined by,

$$\kappa = \left( \frac{Q}{A_T} \right) / \left( \frac{\Delta T}{L_z} \right) \quad (5)$$

where $J_Q = Q/A_T$, they argued that $S_3$ (equation 3) could be estimated by multiplying $S_4$ (equation 4) by $\kappa$. This was based on using a simple linear model that assumed that the thermal conductivity of the substrate and magnetic film were well matched.

For samples where this is not the case (the thermal conductivity is not well matched), we showed that the substrate can play a significant role in determining the value of $\Delta T$ across the active material – the magnetic layer – in a thin film device [19]. This highlights a significant disadvantage of using the spin Seebeck coefficients outlined in equations (1)-(3) for thin film devices, where the measurement of $\Delta T$ is a poor indicator of the temperature gradient across the active material. To circumvent this problem, we argued that in the equilibrium condition a simple thermal model can be used to determine the heat flux through the entire sample,

$$Q = \frac{A_T \Delta T}{\left( \frac{d_1}{\kappa_1} \right) + \left( \frac{d_2}{\kappa_2} \right) + \left( \frac{d_3}{\kappa_3} \right)} \quad (6)$$

where $\{d_1, d_2, d_3\}$ and $\{\kappa_1, \kappa_2, \kappa_3\}$ are the thicknesses and thermal conductivities of the top layer (1), FM layer (2) and substrate (3), respectively; and $\Delta T$ is the temperature difference across the entire device. It can be seen from this, that the temperature gradient is a function of the thermal conductivities of each layer and that the temperature difference across the 'active' magnetic layer can be thermally shunted by substrates with relatively low thermal conductivities (for example, see the comparison made between SrTiO$_3$ and glass Fe$_3$O$_4$:Pt in [19]). Note that this treatment of the thermal profile is, an oversimplification as it does not take into account any temperature drops at interfaces.

Given that $J_Q$ will be constant across the sample once thermal equilibrium has been reached, and that it will be proportional to the temperature gradient across the active magnetic layer, of thickness $d_2$, we can write the thermal gradient as,

$$\nabla T = \frac{\Delta T}{d_2} = \frac{J_Q}{d_2}\left(\frac{d_2}{\kappa_2}\right) = \frac{J_Q}{\kappa_2} \quad (7)$$

where for the magnetic layer $L_z = d_2$ and we now have a direct relationship between the thermal gradient, $\nabla T$, and the heat flux $J_Q$ in the active material.

Finally, whilst there has been limited discussion of the scaling of the spin Seebeck effect with area, it was argued by Kirihara *et al.*, that the spin Seebeck effect may also scale with area [27] based on the change in internal resistance of the paramagnetic layer,

$$R_0 = \frac{\rho L_y}{L_x t_{Pt}} \quad (8)$$

where $\rho$ is the resistivity, $t_{Pt}$ is the thickness, and $L_x$ is the width of the Pt layer. This suggests that the maximum power extracted is,

$$P_{max} \propto \frac{V^2}{R_0} \propto L_x L_y \quad (9)$$

Note that this observation would only indicate an increase in power output of the bilayer (not the voltage).

## IV. RESULTS AND DISCUSSION

To test normalization of the spin Seebeck measurements for both heat flux and sample area, we prepared several Fe$_3$O$_4$:Pt samples on glass substrates of varying thickness, where the Pt thickness was kept constant at 5 nm ($\pm$0.5 nm) in order to minimize any variation in the observed voltage, $V_{ISHE}$, due to the inverse spin Hall effect [19]. $V_{ISHE}$ was then measured for various thermal gradients as a function of applied magnetic field.

We first took one of the Fe$_3$O$_4$:Pt samples from our study (22x22x0.5 mm glass substrate, 80 nm Fe$_3$O$_4$, 5 nm Pt) and measured it in various orientations, as summarized in Figure 2(a)&(e).

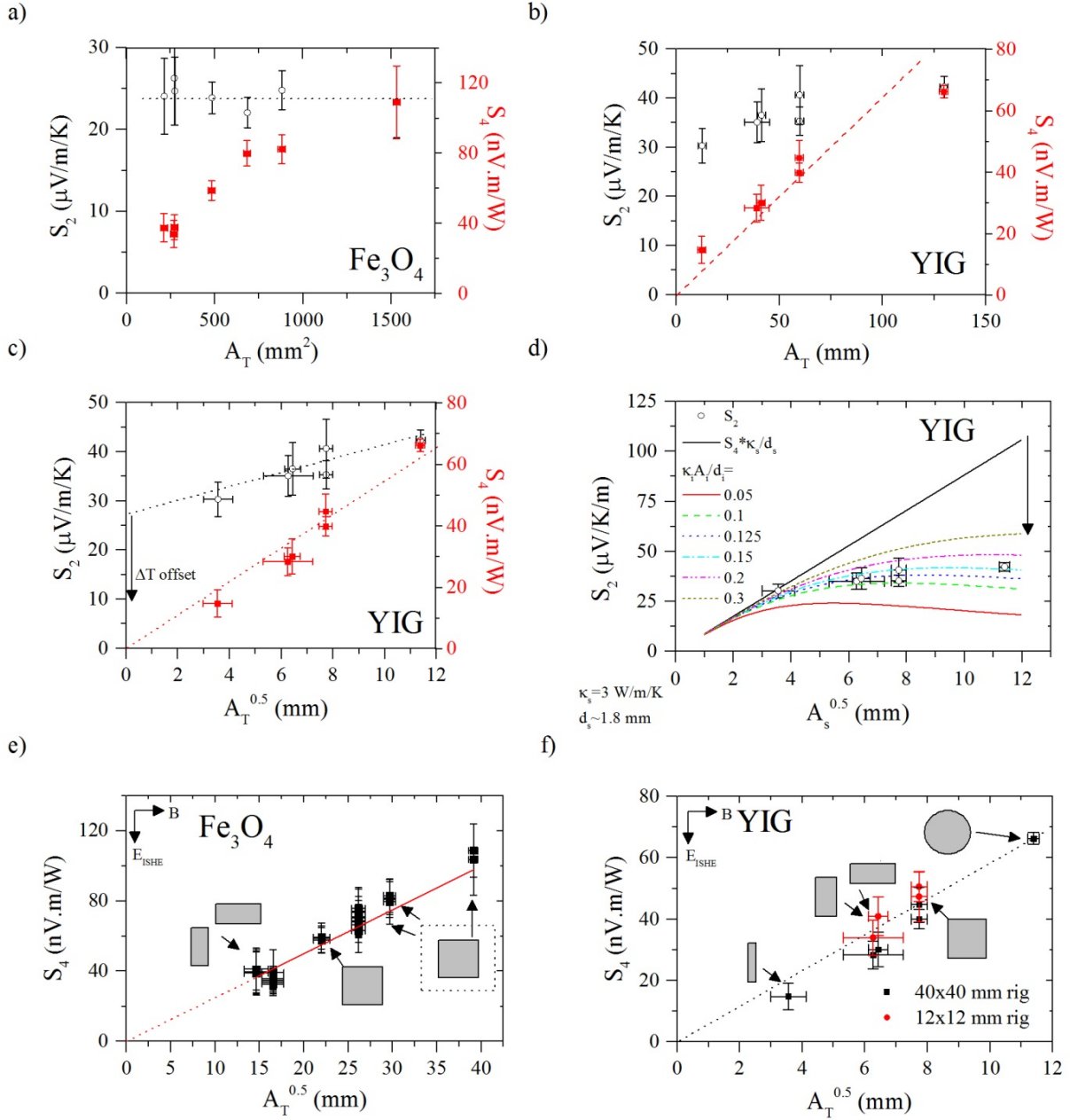

Figure 2 Summary of normalization measurements for thin film $Fe_3O_4$:Pt and bulk polycrystalline YIG:Pt. (a)&(b) $S_2$ (open symbols) and $S_4$ (closed symbols) as a function of $A_T$ (as $L_z$ is fixed $S_2$ and $S_3$ are equivalent). The lines are guides for the eye. (c) $S_2$ (open symbols) and $S_4$ (closed symbols) as a function of $A_T^{0.5}$ for the YIG sample. (d) Simulated change in the measurement of $S_2$ as a function of $A_T^{0.5}$ when the interfacial thermal resistance is varied. (e) & (f) $S_4$ as a function of $A_T^{0.5}$ for the thin film and bulk sample, respectively. Inset sketches indicate aspect ratio for the individual data points. For the thin film measurements a buffer layer (same thickness glass substrate) was inserted in order to increase $A_T$; as indicated by the inset sketch with dotted outline. Where x-axis errors are not visible they are smaller than the symbol used.

To increase the area of the measured sample, a buffer layer between the two Peltier cells was introduced (substrate of same thickness as the sample – on the assumption that control of $J_Q$ by the magnetic layer is largely negligible). This effectively increased the thermal contact area $A_T$, by increasing $\{L^T_x, L^T_y\}$ whilst $\{L_x, L_y, L_z\}$ were fixed (see Figure 1(d)). Secondly, the 22x22 mm sample was cleaved in two and measured along the long and short sides, thus changing $A_T$, the aspect ratio and $L_y$. Errors were determined from the combination of uncertainty in contact separation $L_y$, thermal contact area $A_T$, and the noise floor of the $V_{ISHE}$ voltage.

Figure 2(a) shows the calculated spin Seebeck coefficients $S_2$ (left) and $S_4$ (right axis) as the area, $A_T$, was increased. Initially this data suggests that there appears to be an increase in the coefficient measured by the heat flux method ($S_4$), but not by the temperature difference method ($S_2$). In other words, for the same sample, the aspect ratio and total area seem to have an impact on the determination of $S_4$. In addition, by plotting $S_4$ against various combinations of $L^T_x$, $L^T_y$, and $A_T$ (as shown in Appendix A, Figure A1), we found that the most likely way to reduce the measurement to a constant value was to divide through by $A_T^{0.5}$. This suggests that the data requires the following normalization relation in order to produce a geometry independent coefficient, $S_5$,

$$S_5 = \frac{V_{ISHE}}{L_y\left(\frac{Q}{\sqrt{A_T}}\right)} \quad (10)$$

This is further demonstrated by the linear dependence observed for the thin film data in Figure 2(e). Note that for this series of measurements, as $A_T$ exceeded 484 mm² the area of active material was no longer increasing (as $A_T$ was further increased by introducing a buffer layer with matched thermal conductivity). So whilst the heat flux across the active material was constant (by definition of $S_4$ in equation (4)) there was an apparent increase in the voltage per unit heat flux. It could be argued that this is simply due to heat losses, however, as will be shown later, this trend was still observed for samples where $A_T$ was varied and measured in a setup with matching Peltier surface area.

Given that in the steady state, $S_4\kappa=S_2$ (equations 2-7), the mismatch between the values of $S_2$ and $S_4$ determined for the thin film samples indicates that there is a measurement artefact that needs to be resolved. To test for this, we also measured bulk YIG samples as a function of $A_T$, where the temperature difference across the active material would now be an order of magnitude larger (than our thin films) and thus, less prone to errors such as interfacial thermal resistance. In addition, due to the insulating nature of YIG, any contribution to $V_{ISHE}$ due to the anomalous Nernst effect (ANE) will no longer be present.

Figure 2(b), (c) & (f), shows the same normalization measurements for the bulk YIG:Pt sample as a function of $A_T$. Here, the sample was measured in both the 12x12 mm and 40x40 mm sample holders, and cut from the original 13 mm pellet to various sizes. Notice that the data from the 12x12 mm and 40x40 mm measurement set-ups are the same within error, and that it still indicates possible scaling with $A_T^{0.5}$. For this dataset, plotting $S_4$ as a function of $L_x^T$ also indicated a linear trend (as seen in Appendix A, Figure A2), but this is for the case where $A_T=L_xL_y$ (as $L_x^T=L_x$ and $L_y^T = L_y$), i.e. it does not account for non-standard geometries, where the contacts are not necessarily at the edges of the sample. In either case, there is still a pronounced increase in the measured voltage as the sample size increases that cannot be resolved by normalizing simply by contact separation and/or resistance.

For the bulk samples, however, there is now an indication of scaling of the temperature difference method ($S_2$) with $A_T$. This can be seen in Figure 2(b) and (c) where $S_2$ is plotted alongside $S_4$ as a function of $A_T$ and $A_T^{0.5}$. The difference between these measurements and that of the thin films is firstly that the measurement of $\Delta T$ across the (active) magnetic sample is now direct, and secondly, that the increased thickness of the sample (1.8 mm rather than 0.15 – 0.9 mm) limits the impact of temperature drops at the Peltier:sample interface due to thermal resistance. We argue that these are the reasons why the increase of $S_2$ with $A_T$ is now obvious for the bulk samples.

In order to demonstrate this, in Figure 2(d), we present a simple model of the expected trend for $S_2$, as the sample area is increased. We first rewrite equation (2) to include the systematic error in measurement of $\Delta T$,

$$S_2' = \frac{V_{ISHE}}{L_y(\Delta T_s+\Delta T_i)} \quad (11)$$

where $\Delta T_s$ is the actual temperature drop across the sample and $\Delta T_i$ is the temperature drop at the sample:thermocouple interface(s). (For the true measurement of $S_2$, the temperature offset, $\Delta T_i$, should be subtracted.) We then make the assumption that $\Delta T_i$ will be approximately constant (for the same $Q$). We argue that if the individual measurements are well controlled (i.e. similar sample mounting, use of thermal grease and comparable force when clamping the sample between the two Peltiers), then this is a reasonable assumption as it is likely driven by the thermal properties of the interface to which the thermocouples are attached (i.e. an effective thermal conductance). If this were not the case, then there would be considerably more scatter in the data for measurement of $S_2$, both here and in the literature.

In this case, according to equation (5), we can define the heat, $Q$, passing through the sample and the interface as follows,

$$Q = \frac{\Delta T^i \kappa^i}{d^i} A^i \quad (12)$$
$$Q = \frac{\Delta T^s \kappa^s}{d^s} A^s \quad (13)$$

where $\kappa_i$ & $\kappa_s$ are the thermal conductivities, $A_i$ & $A_s$ are the cross sectional areas, and $d_i$ & $d_s$ are the thicknesses of the interface and the sample, respectively. If we combine equations (12) & (13), we can write $\Delta T_i$ in terms of $\Delta T_s$ as,

$$\Delta T_i = \Delta T_s \left(\frac{d_i \kappa_s A_s}{d_s \kappa_i A_i}\right) \quad (14)$$

and finally the measured temperature difference, $\Delta T$, as,

$$\Delta T = \Delta T_s + \Delta T_i = \Delta T_s \left(1 + \frac{d_i \kappa_s A_s}{d_s \kappa_i A_i}\right) \quad (15)$$

From equation (15) it should be clear that as the sample area is increased, the influence of the interfacial resistance (for a given measurement setup, where $\kappa_i A_i/d_i$ remains approximately constant) increases. Similarly, for the thin film samples, as $d_s$ is of the order of 80 nm, the impact of $\Delta T_i$ will be more pronounced. This was shown previously, where $\Delta T_s$ was found to be 0.01% of the total measured $\Delta T$ in our thin film Fe$_3$O$_4$. [19]

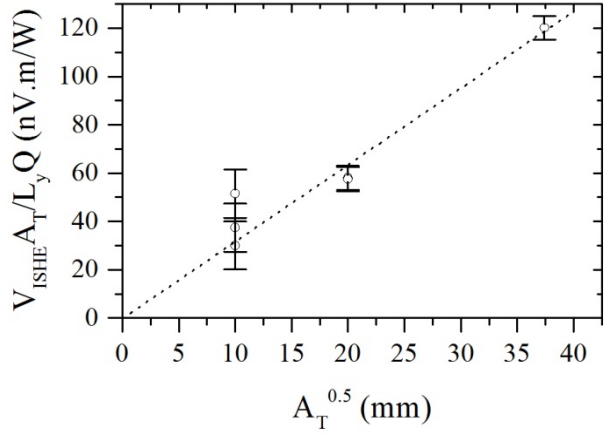

| | 10x10 | 20x20 | 40x40 |
|---|---|---|---|
| $S_p$ (top) | 0.11 | 0.15 | 0.14 |
| $A_T$ (mm$^2$) | 100 | 400 | 1400 |
| $L_y$ (mm) | 7 | 20 | 38 |
| $I_{resistor}$ (A) | 0.24 | 0.159 | 0.079 |
| $R$ (Ω) | 54 | 57 | 197 |
| $Q_{resistor}$ (W) | 3.11 | 1.44 | 1.23 |
| $V_p$ (V) | 0.079 | 0.186 | 0.127 |
| $Q_{top}$ (W) | 0.72 | 1.28 | 0.91 |
| $V_{ISHE}$ (µV) | 1.85 | 3.84 | 2.91 |
| $V_{ISHE}/Q_{top}$ (µV/W) | 2.57 | 3 | 3.2 |
| $V_{ISHE}/Q_{top}L_y$ (µV/m/W) | 367.1 | 150 | 84 |
| $V_{ISHE}A_T/Q_{top}L_y$ (nV.m/W) | 36.71 | 60 | 117.6 |
| $V_{ISHE}/Q_{resistor}$ (µV/W) | 0.59 | 2.67 | 2.37 |
| $V_{ISHE}A_T/Q_{resistor}L_y$ (nV.m/W) | 8.43 | 53.3 | 87 |

Figure 3 - Spin Seebeck measurement(s) of Fe$_3$O$_4$:Pt 50x50 mm thin film which was cleaved to match measurement areas of 10x10, 20x20, and 40x40mm Peltier cells. The line is a guide for the eye and x-error was smaller than the symbols used. The table shows a set of values used to obtain this plot where $S_p$ is the Peltier calibration coefficient, $I_{resistor}$, the current supplied to a resistor, $R$, mounted onto the top surface of the top Peltier cell. $Q_{resistor}$ is the resulting heat source due to the resistor, $V_p$ is the measured voltage for the top Peltier cell, and $Q_{top}$ the measured heat from the top Peltier (=$V_p/S_p$).

Equation (15) was used to simulate $S_2'$, i.e. how the temperature dependent spin Seebeck coefficient ($S_2$) is modified as a result of $\Delta T^i$. In this case we used the measured values of $S_4$, and the known thermal conductivity and thickness of the sample (3 W/K/m and 1.8 mm) to estimate $S_2$ where $\Delta T_i$ was negligible. We then multiplied this by $\Delta T_s/(\Delta T_s + \Delta T_i)$ to obtain $S_2'$ for various values of $\kappa_{eff}$ (=$\kappa_i A_i/d_i$). Note that as $A_T$ is increased, $S_2'$ appears to saturate, and that this occurs sooner for higher values of $\kappa_{eff}$ (as seen in Figure 2(d)).

To summarize, there is competition between an increase in $S_2$ as $A_T$ increases (which is observed with the heat flux method), and a decrease in $S_2$ due to $\Delta T_i$. As stated earlier, the advantage of defining a spin Seebeck coefficient dependent on $J_Q$ rather than $\Delta T$ is that it will be independent of thermal contact resistance [16], and the substrate.

Finally, to confirm that the observed trend is not a result of heat losses, when the Peltier area was not matched to the sample area, we repeated the thin film measurement for 3 separate measurement set-ups, where the Peltier area was 10x10mm, 20x20mm, and 40x40mm and a single film was cleaved to match this. In this case, the temperature gradient was driven by a resistor (R) mounted onto the top surface of the top Peltier so that heat flow could be monitored either side of the sample. The sample used here was deposited onto a 50x50 mm substrate, where we might expect a thickness variation of the Pt layer of approximately 10% across the sample area. To quantify the impact of this thickness variation on the measurement, we took 3 10x10 mm pieces from corners of the sample, to measure the scatter in the spin Seebeck coefficient. This can be seen in Figure 3, where for one of the 10x10 mm samples $S_4$ = 51.63 nV.m/K compared to 30.1 and 37.43 nV.m/K, and gives an upper limit of the expected variation in the film (see supplementary information for more information). The result of these measurements is given alongside a tabulated example of the key figures for a subset of the measurements (i.e. one heat flux for each area) in Figure 3. Notice here, that the difference in $S_4$ is greater than a factor of 3 between the smallest and largest sample(s). Even when scaling by the power supplied to the resistive heater (where we expect heat losses), this increase in $S_4$ is observed.

In addition, we also performed measurements on a sample with fixed area, $A_T$, but modifying $L_x$ and $L_y$ by scribing away areas of the sample (so that it can be considered as thermally connected but electrically isolated). In this case, $V_{ISHE}$ always scaled with $L_y$, as typically expected for a measurement where $A_T$ does not change.

To summarize, we observed, for fixed thickness of the magnetic and Pt layers:

1. At constant $A_T$ and $J_Q$: Linear scaling of $V_{ISHE}$ with $L_y$, which is independent of sample geometry (as expected).
2. At constant $A_T$ and $L_y$: Linear scaling of $V_{ISHE}$ with $\Delta T$ and $J_Q$ (as expected).
3. At constant $A_T$ and $J_Q$: No observable change, when areas of the sample had been rendered inactive by scribing a line between the magnetic layers (i.e. scribing away the electric contact and modifying the resistance $R$ and charge current in that section, $I_c$).

4. At constant $J_Q$: An increase in the spin Seebeck coefficient, $S_4$ $(=V_{ISHE}A_T/L_yQ)$ when $A_T$ was increased.
5. At constant $J_Q$ and constant $L_y$: An increase in the spin Seebeck coefficient, $S_4$ $(=V_{ISHE}A_T/L_yQ)$ when $A_T$ was increased.

The above observations indicate that there is an additional parameter that is being overlooked when evaluating the magnitude of voltage generation due to the inverse spin Hall effect, or thermal pumping of spin current, $I_s$, into the Pt layer. At this stage it is not clear what this mechanism is, but we speculate that it could be due to the assumption that the spin current density, $J_S$, injected in the Pt layer at the Ferromagnet:Paramagnet interface is directly proportional to $J_Q$. Finally, this suggests that $S_5$, as defined in Equation (10), is the most appropriate coefficient to use when comparing different bilayer systems, with arbitrary area, $A_T$.

Overall, for the thin film $Fe_3O_4$ samples we can compare our results to that of Ramos *et al.*,[12]. In this work they reported $S_2 = 150$ µV/K/m for 50 nm $Fe_3O_4$ deposited epitaxially onto a 0.5 mm thick, 8x4 mm $SrTiO_3$ substrate, with a 5 nm Pt layer. If we assume that the temperature gradient is controlled by the thermal conductivity of the substrate so that $S_4 = S_2\kappa/d$ ($k_{SrTiO3}$=11.9 W/m/K for $SrTiO_3$) this would give $S_4 = 6.3$ nV.m/W. For the 0.5 mm thick $SiO_2$ substrate sample in Figure 2 of this work (80 nm $Fe_3O_4$, 5 nm Pt, 10x10 mm) applying a similar approach ($k_{SiO2}$~1 W/m/K, $d$ = 0.5 mm) would give $S_4 \sim 12$ nV.m/W, where we measure approximately 30 nV.m/W by the heat flux method. This data shows that for these $Fe_3O_4$:Pt bilayers $S_5 = 3\pm0.2$ µV $W^{-1}$. The difference between the values obtained by the heat flux and temperature difference methods are not unexpected as Sola *et al.*[17] showed that the $\Delta T$ method can routinely underestimate the spin Seebeck coefficient. However, it does demonstrate the comparable quality of our thin films.

For bulk YIG measurements we obtain $S_5 = S_4/A_T^{0.5} = 4.8\pm0.34$ µV/W and $S_2/A_T^{0.5} = 8\pm0.24$ mV/K/$m^2$ (using $\kappa = 3$ W/m/K and $d = 1.8$ mm). These values are reasonable given the porosity of this YIG pellet. For comparison, measurements of bulk polycrystalline YIG by Saiga *et al.*, [28] found $S_1$ of up to 5 µV/K, when annealing to improve the interface equality. Given the dimensions $L_x = 5$ mm, $L_y = 2$ mm, $L_z = 1$ mm, this amounts to $S_2 = 1000$ µV/K/m and $S_2/A_T^{0.5} = 316$ mV/K/$m^2$, or $S_5 = 45$ µV/W (assuming $\kappa = 7$ W/m/K). If we compared to direct heat flux measurements by Sola *et al.*, with thin film YIG on GGG [17], they measured $S_4 = 0.11$ µV.m/W for a 5x2 mm sample. Hence, $S_5 = 34$ µV/W. Whilst these values are an order of magnitude larger than our measurement, for these examples care was taken to maximize the interface quality. In addition, the Pt thickness was smaller ($t_{Pt} < 3$ nm), which would be expected to increase $V_{ISHE}$ further.

To further demonstrate the impact of thermal resistance on the measurement of $S_2$, we show in Figure 4, results of spin Seebeck measurements for our $Fe_3O_4$:Pt bilayers as a function of substrate and $Fe_3O_4$ thickness ($t_{substrate}$ and $t_{Fe3O4}$ respectively). Note that there was a scatter in our individual data points for $S_2$ that can be attributed to varying thermal resistance between repeated measurements for the same sample (see supplementary information for individual measurements); this led to larger error bars. In Figure 4(a) we show a summary of all measurements as a function of substrate thickness. As can be seen, $S_2$ decreased linearly with increasing thickness, as a larger proportion of $\Delta T$ was shunted across the substrate. This is not unexpected. As soon as the data was normalized according to Equation (10), as is also shown in Figure 4(a) we obtained a relatively constant value (within error, and due to variations that might be expected from changing interface quality [14] or small differences in $t_{Pt}$ and the resistivity).

With regards to the thickness dependence of the spin Seebeck effect (Figure 4(b)) we also see collapse of the data onto the expected saturation at around 80 nm. Again, this is not unusual: saturation of the signal would be expected as the thickness of the $Fe_3O_4$ layer increases, given that the volume magnetization also saturates at around 80 nm (see supplementary information, Fig. S3). The dashed line in this figure shows the corresponding fit to this data using the approach of Anadón *et al.*,[25] who found a magnon accumulation length of 17 (± 3) nm at 300K from this fit. The slight decrease in the spin Seebeck coefficient for the thickest sample could be due to a change in the morphology of the thicker film and is the focus of current work.

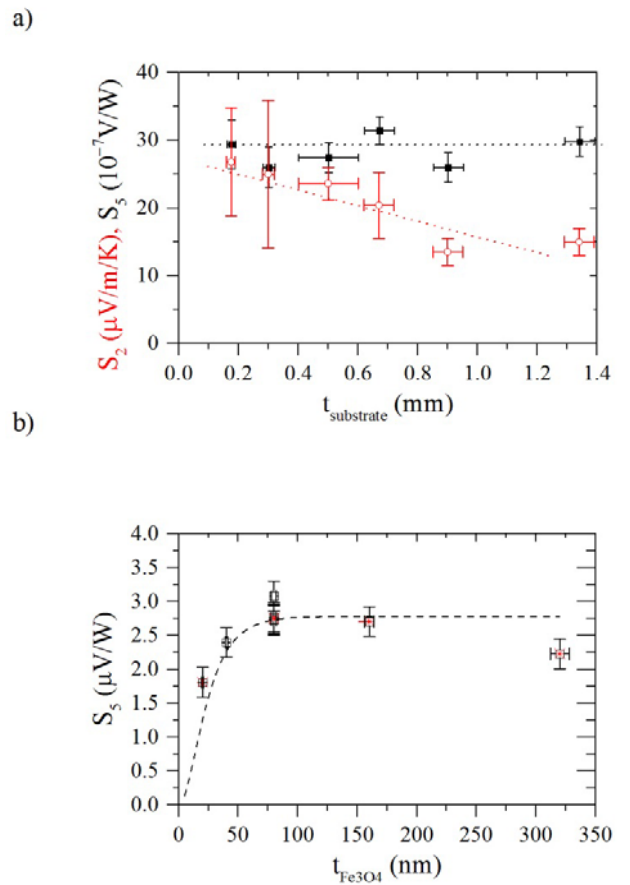

Figure 4 Summary of measurements as a function of substrate and $Fe_3O_4$ thickness, $t_{substrate}$ and $t_{Fe3O4}$, respectively. (a) Average value for $S_2$ (open symbols) and $S_5$ $(=V_{ISHE}A_T^{0.5}/L_yQ)$ (closed symbols) as a function of the substrate thickness, $t_{substrate}$. (d) $S_5$ as a function of $t_{Fe3O4}$. Dashed line shows a fit according to linear response theory[25], with a magnon accumulation length of 17 nm.

## V. CONCLUSION

To summarize, we have demonstrated that the heat flux method is suitable for obtaining substrate independent measurements of the spin Seebeck effect, where the thermal conductivity of the thin film need not be known, and have proposed an alternative spin Seebeck coefficient ($S_5$) for comparison of different material systems. By measuring the spin Seebeck effect as the thickness of the substrate, $Fe_3O_4$ layer and sample dimensions $\{L_x, L_y, A_T\}$ were varied, we demonstrated that for the same material system (assuming $t_{Pt}$ is constant at 5 nm) this method reliably returned a value of 3.0±0.2 µV W$^{-1}$. Since this result holds for different values of $t_{substrate}$, the heat flux normalization method could, therefore, be used to compare similar samples where the substrate thickness or type were varied. This would thus be a useful metric for comparison of prospective material systems (for application).

Lastly, we demonstrated unexpected scaling of the spin Seebeck coefficient (as typically defined in the literature). These results indicate that the voltage is proportional to the available energy per unit length ($Q$), which we speculate could be a result of thermal spin injection into the paramagnetic layer before detection. The exact nature of this scaling will be the topic of future studies.

## ACKNOWLEDGEMENTS
This work was supported by EPSRC First Grant (EP/L024918/1) and Fellowship (EP/P006221/1) and the Loughborough School of Science Strategic Major Operational Fund. KM would also like to thank M.D. Cropper for their help with use of the PLD system for sample fabrication and M Greenaway and J Betouras for useful discussions. Supporting data will be made available via the Loughborough data repository under doi 10.17028/rd.lboro.5117578.

## APPENDIX A: ADDITIONAL PLOTS OF $S_4$ FOR THE THIN FILM AND BULK SAMPLES

Figures A1 and A2 show additional plots of the data presented in Figure 2 (e) & (f), as $A_T$ was varied. This includes parameters such as the horizontal and vertical lengths, $L_x^T$, $L_y^T$, their ratio, and the area $A_T$. The dashed lines are guides for the eye.

## APPENDIX B: HEAT LOSS CONSIDERATIONS

In an ideal case, the sample area would be chosen such that it matches the area of the hot and cold bath(s) in the measurement. Given that this paper studies the impact of sample size (where $A_T$ is varied), it could be argued that for samples where $A_T$ is less than the area of the top and bottom Peltier cells, there are significant thermal losses (radiation and convection). This was monitored during measurement by plotting the change in the measured $J_Q$ as a function of $\Delta T$ (see Figure 1(e)). It was also confirmed by obtaining comparable measurements in air and under vacuum, as detailed in the supplementary information.

We consider here the potential impact of radiative losses, as defined by:

$$Q = \varepsilon \sigma A (T_h^4 - T_c^4) \quad (16)$$

where $\varepsilon$ is the emissivity of the radiating surface ($\varepsilon = 1$ for a perfect radiator), $\sigma$ is the Stefan-Boltzman constant ($\sigma = 5.67\times10^{-8}$ W/m$^2$/K$^4$), $A$ is the radiating surface area, and $T_h$ and $T_c$ are the temperatures of the hot and cold (ambient) surfaces respectively. For the sample environment used in these measurements there will be 3 sources of radiative heat loss:

1) Top surface of the Peltier cell where the sample is not connected (forming a conductive path).
2) Bottom surface of the Peltier cell where sample is not connected.
3) The sides of the sample.

Sources 1&2 can be considered simultaneously given that $A$ will be the same, and it is reasonable to assume that the majority of the heat lost from the top surface will be measured by the bottom surface (i.e. overestimating $J_Q$). Thus for 1&2:

$$Q_{rad} = \varepsilon \sigma (A_P - A_T)(T_h^4 - T_c^4) \quad (17)$$

where $A_p$ is the Peltier area.

For the worst case scenario for radiative losses 1&2 the smallest samples measured in the 40x40mm$^2$ Peltier setup had $A_T$=100 mm$^2$. In this case, $A_P$-$A_T$=1500 mm$^2$. Inserting this into equation (17), the radiative heat loss would be $Q$ = 0.045 W for a measured heat flow of 2.5 W (1.8%).

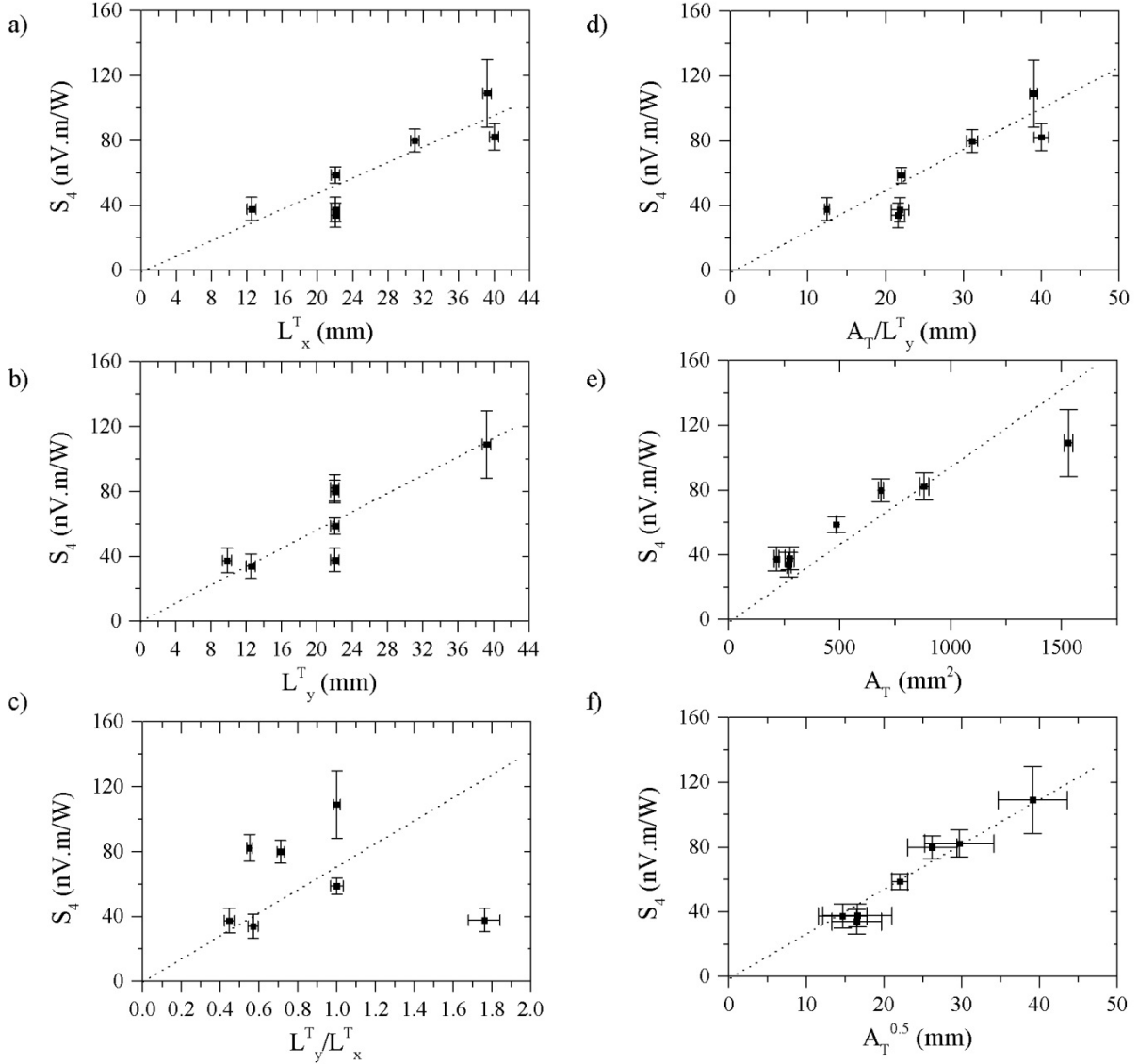

Figure A1 (a) – (f) Spin Seebeck measurement(s) for the $Fe_3O_4$:Pt thin film as $A_T$ ($=L^T_x \cdot L^T_y$), $L_x$ and $L_y$ were varied, plotted against $L^T_x$, $L^T_y$, $L^T_y/L^T_x$, $A_T/L^T_y$, $A_T$ and $A_T^{0.5}$. Where x-axis errors are not visible they are smaller than the symbol used.

The worst case scenario for radiative loss via the sides of the sample was for the thickest samples, with largest area. The thickness of the measured thin films were 0.17-1.34 mm, with a maximum radiative area (from the sides) of 171.2 mm². For a typical temperature difference of 5 K this would amount to radiative loss of approximately 0.0045 W (0.2% of $Q$). For the YIG samples, with thickness = 1.8 mm, and temperature differences of up to 25 K the radiative loss would be 0.012 W (0.5% of $Q$). In this context, for these temperature differences, radiative loss is negligible with respect to the changes seen as a function of $A_T^{0.5}$.

As a general rule of thumb, radiative heat losses in this system can be quickly assessed by monitoring $J_Q$ vs. $\Delta T$. Given equations (5) and (16), for the ideal case the heat flux through the sample should be linear with respect to $\Delta T$. As soon as the radiative heat losses become significant with respect to the heat flow through the sample, this linearity breaks down as a larger $J_Q$ will be observed per unit $\Delta T$.

Conversely, for much thicker samples, where radiative loss from the sides of the sample starts to become significant, such as is the case with the bulk YIG samples, heat loss would result in a lower measured $J_Q$ per unit $\Delta T$ as well as non-linearity of $J_Q(\Delta T)$. We only present data for values of $\Delta T$ where this linearity was preserved.

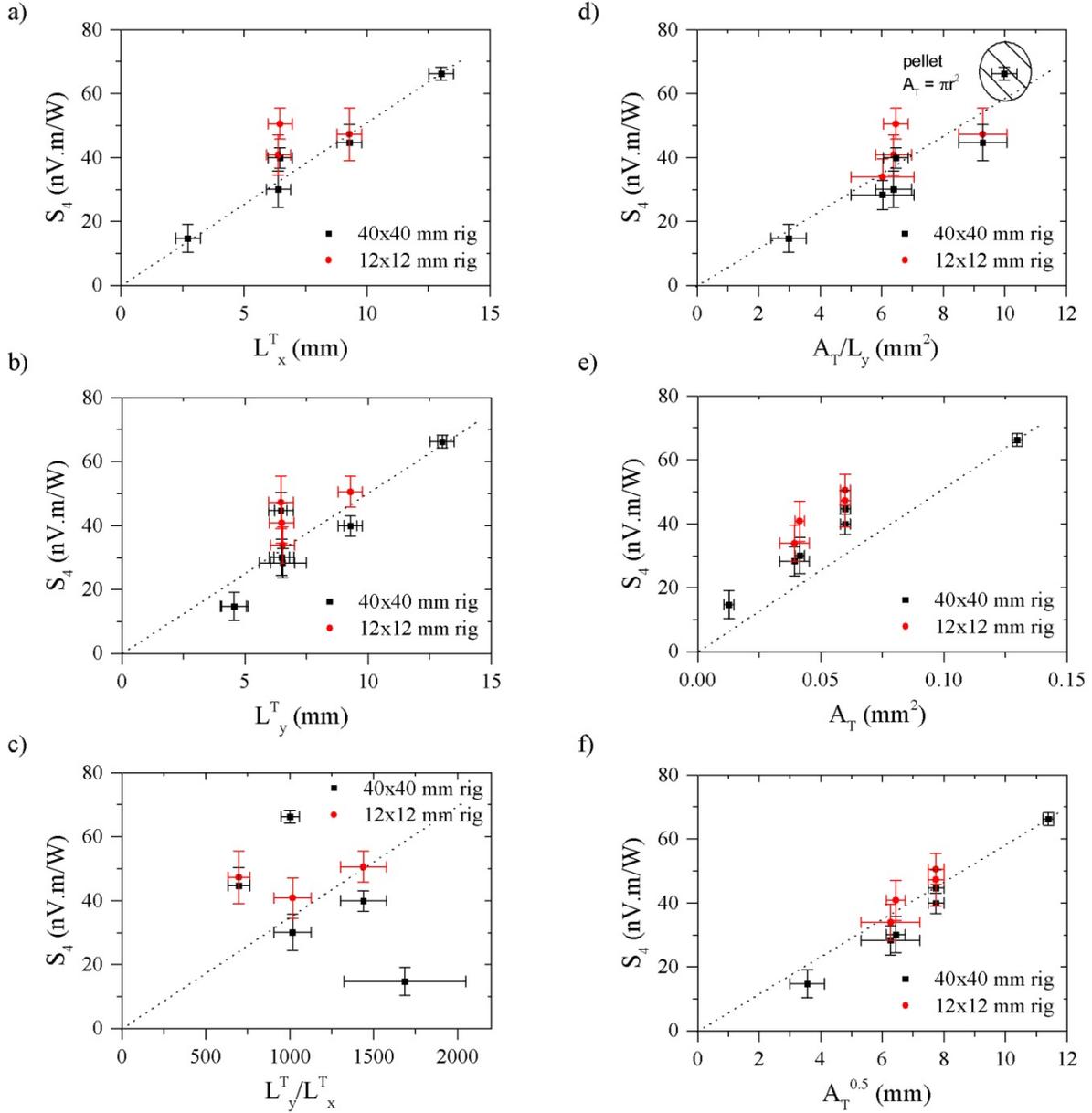

Figure A2 (a) – (f) Spin Seebeck measurement(s) of YIG:Pt as $A_T$ (=$L^T_x L^T_y$ except where sample was 13 mm pellet), $L^T_x$ and $L^T_y$ were varied, plotted against $L^T_x$, $L^T_y$, $L^T_y/L^T_x$, $A_T/L_y$, $A_T$ and $A_T^{0.5}$.

# Supplementary information for "Scaling of the spin Seebeck effect in bulk and thin film"

**K. Morrison[1], A J Caruana[1,2], C. Cox[1], T.A Rose[1]**

[1] Department of Physics, Loughborough University, Loughborough, LE11 3TU, United Kingdom
[2] ISIS Neutron and Muon Source, Didcot, Oxfordshire, OX11 0QX

**1 Introduction** Extensive experimental details and additional characterization of the films presented in "Scaling of the spin Seebeck effect in bulk and thin film" is given in this supplementary information.

Additional X-ray diffraction (XRD), resistivity and X-ray reflectivity (XRR) data is used to demonstrate the quality of $Fe_3O_4$ thin films – exhibiting the characteristic Verwey transition for this phase[1], as well as the expected magnetic characteristics as a function of temperature. This is followed by additional XRD, and XRR data for the $Fe_3O_4$:Pt bilayers to further demonstrate the magnetic and structural quality of the films, as well as XRD and SEM data of the YIG pellet chosen for this study. Finally, further details of the spin Seebeck measurements are given, alongside the additional 'scribing' study data.

## 2 Experimental Details
### 2.1 Sample preparation

The series of films prepared for this study are listed in Table S1. They were prepared in vacuum (base pressure of $5\times10^{-9}$ mbar) by pulsed laser deposition (PLD) using a frequency doubled Nd:YAG laser (Quanta Ray GCR-5) with a wavelength of 532 nm and 10 Hz repetition rate. The films were deposited onto 10 x 10 mm and 22 x 22 mm glass slides (Agar Scientific borosilicate coverglasses), or 24 x 32 and 50 x 50 mm fused silica, which were baked out at 400 °C prior to the growth of the $Fe_3O_4$ layer at the same substrate temperature. The samples were then left to cool in vacuum until they reached room temperature, at which point the Pt layer was deposited. The $Fe_3O_4$ and Pt layers were deposited from stoichiometric $Fe_2O_3$ (Pi-Kem purity 99.9%) and elemental Pt (Testbourne purity 99.99%) targets using ablation fluences of $1.9\pm0.1$ and $3.7\pm0.2$ J cm$^{-2}$ respectively. The target to substrate distance was 110 mm. The Pt layer thickness was kept constant at $5\pm0.5$ nm for all films, whilst the substrate and $Fe_3O_4$ thicknesses were varied.

Bulk YIG was prepared by the solid state method. Stoichiometric amounts of $Y_2O_3$ and $Fe_2O_3$ starting powders (Sigma Aldrich 99.999% and 99.995% trace metals basis, respectively) were ground and mixed together before calcining in air at 1050 °C for 24 hours. Approximately 0.5g of the calcined powder was then dry pressed into a 13 mm diameter, $1.8\pm0.2$ mm thick cylindrical pellet. The pellet was then sintered at 1400 °C for 12 hours, after which, it was checked by XRD prior to sputtering 5 nm of Pt onto the as prepared surface using a benchtop Quorum turbo-pumped sputter coater. Samples were cut to size thereafter, using an IsoMet low speed precision cutter.

### 2.2 Structural characterization
XRD was obtained using a Bruker D2 Phaser in the standard Bragg-Brentano geometry using Cu K$_\alpha$ radiation; a Ni filter was used to remove the Cu K$_\beta$ line. The samples were rotated at a constant 15 rpm throughout the scan to increase the number of crystallites being sampled. XRR measurements were taken using a modified Siemens D5000 diffractometer comprising a graphite monochromator reflecting only Cu K$_\alpha$ into the detector. The incident beam was collimated with a 0.05 mm divergence slit, while the reflected beam was passed through a 0.2 mm anti-scatter slit followed by a 0.1 mm resolution slit. XRR data was fitted using the GenX software package [2].

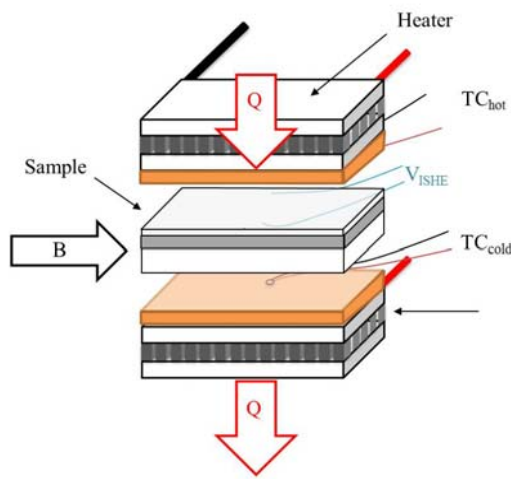

**FIG. S1** Schematic of the spin Seebeck measurement system. Top and bottom Peltier cells act as heat source and heat flux monitor, respectively. Copper plates on the sample facing side of the Peltier cells provide high thermal conductivity at the interface, thus reducing lateral temperature gradients. GE varnish and mica are used to electrically isolate the copper surface from voltage contacts. Thermocouples monitor the temperature (with respect to the cold sink) at the top and bottom of the sample. Finally, the voltage, $V_{ISHE}$ is monitored as Q and B are varied. Note that the thermal contact area, $A_T$, is defined as the area of sample connecting the top and bottom Peltier cells.

**2.3 Magnetic and electric characterization**
Magnetization as a function of applied magnetic field was measured using a Quantum Design Magnetic Properties Measurement System (SQUID). Several bilayer samples were characterized as a function of field at room temperature, whilst the Fe$_3$O$_4$ film (without Pt contact) was measured at 5 K, 100-130 K, and 295 K in order to observe the characteristic magnetic Verwey transition at 115 K.

Room temperature sheet resistance was obtained using the Van der Pauw method with a Keithley 6221/2182 nanovoltmeter and current source in conjunction with a Keithley 705 scanner.

**2.4 Thermoelectric characterization**
Spin Seebeck measurements were obtained from a set-up similar to that of Sola *et al.* [3], and outlined in Figure S1. There were 4 separate set-ups, optimized for 10x10, 12x12, 20x20 and 40x40 mm samples. The majority of data presented in the main manuscript were obtained using the 12x12 and 40x40mm setups unless otherwise stated. The thin film was sandwiched between 2 Peltier cells, where the top Peltier cell (1) acted as a heat source, and the bottom Peltier cell (2) monitored the heat Q, through the sample. A copper sheet was affixed to the surface of the Peltier cells to promote uniform heat transfer.

**Table S1** Summary of the thin film samples measured. The contact separation and corresponding thermal contact area, $A_T$, are also indicated. There were 3 distinct series of sample: (a) where the Fe$_3$O$_4$ and Pt thicknesses, $t_{Fe3O4}$ and $t_{Pt}$ respectively, were kept constant (at a nominal 80 nm:5 nm) and the substrate thickness $t_{substrate}$, was varied, (b) where the Fe$_3$O$_4$ thickness was varied from 20 – 320 nm, whilst the Pt thickness was kept constant, and (c) a large area sample (50x50 mm) that was cleaved into sections for measurement in the 10x10, 20x20, and 40x40 mm setups. Film thickness(es) were determined from fitting XRR data, except where $t_{Fe3O4}$ exceeded 80 nm.

| Sample | $A_T$ (mm$^2$) | $L_y$ (mm) | Substrate | $t_{substrate}$ (mm) | $t_{Pt}$ (nm) | $t_{Fe3O4}$ (nm) |
|---|---|---|---|---|---|---|
| FP1 | 768 | 28.8 | Glass | 0.17 | ~4.8 | 80 |
| FP3 (G6) (FP3 frag) | 136.8 203 | 8.47 12.4 | Glass | 0.3 | 4.9 | 80 |
| FP5 (G34) | 484, 684, 880, 1530 214 270 | 19.3 19.3 7.62 19.44 | Glass | 0.5 | 5.5 | 80 |
| FP6 | 772 | 28.56 | Fused silica | 0.67 0.67+0.67 | 4.8 | 80 |
| FP9 | 983 | 36.66 | Fused silica | 0.9 | 5.4 | 80 |
| F320P (G27) | 66.6 | 6.58 | Glass | 0.3 | 4.4 | 320 |
| F40P (G30) | 100 | 6.26 | Glass | 0.3 | 5.2 | 40 |
| F160P (G31) | 100 | 5.5 | Glass | 0.3 | 4.5 | 160 |
| F20P (G32) | 100 | 7.25 | Glass | 0.3 | 5 | 20 |
| G48 | 2500 1400 400 100 | - 37 20 8.5 | Glass | 0.5 | 5 | 80 |

Two type E thermocouples were mounted on the surface of these copper sheets such that they were in contact with the sample during the measurement. These thermocouples were connected in differential mode to monitor the temperature difference across the sample, $\Delta T$. The Peltier cells were calibrated by comparing the voltage generation, $V_p$, in two scenarios:

(1) where the Peltiers were clamped together, with a resistor attached to the top of one, and the heat flow passing through both was assumed to be constant, such that:

$$Q_{top} = Q_{bottom} \quad (S.1)$$

(2) where the resistor was clamped between the two Peltier cells and the total power measured by the Peltier cells was assumed to be equal to the heat output of the resistor, such that:.

$$Q_{top} + Q_{bottom} = P_{resistor} \quad (S.2)$$

Solving these simultaneous equations produced the sensitivity, $S_p$, which was found to be 0.11 - 0.26 V/W at 300K. The heat flux, $J_Q$, is then determined by:

$$J_Q = V_p/(S_p A_T) \quad (S.3)$$

where $A_T$ is the contact area between the top and bottom Peltier cells. This can be quickly related to the measured temperature difference $\Delta T$, assuming that there are no major thermal losses or temperature drops at the sample:Peltier interface(s):

$$J_Q = k_{eff}\Delta T/L_z \quad (S.4)$$

where $k_{eff}$ is the effective thermal conductivity of the sample.

Given that this paper studies the impact of sample size (thus $A_T$ is varied), it could be argued that for samples where $A_T$ is less than the area of the top and bottom Peltier cells, there are significant thermal losses (radiation and convection). This is monitored by plotting the change in the measured $J_Q$ as a function of $\Delta T$, as was seen in Figure 1. In addition, Figure 6 of the main manuscript (and Fig. S7 here) shows data where the size of the sample was matched to the size of the Peltier cell for measurement of $J_Q$. Finally, Figure S9 here shows similar spin Seebeck measurements in air and vacuum, where both Peltier cells are used to monitor heat flow.

**3 Results**
**3.1 Characterization of the Fe$_3$O$_4$ layer**
Additional characterization data for the Fe$_3$O$_4$ thin films (deposited under identical conditions as the devices measured, but without the Pt top layer) is given in Fig. S2.

The Verwey transition is a characteristic feature of the Fe$_3$O$_4$ phase of iron oxide and manifests as a sharp increase in resistivity and hysteresis below the Verwey transition temperature (typically between 115 and 120 K)[1]. This feature is clearly seen in Fig. S2 (a), where magnetometry shows a sudden increase in hysteresis below 120 K.

In addition, XRD indicates a preference for <111> texture that is moderately sensitive to the direction of the plume during PLD (see repeated film depositions of Fig. S2 (c)). Due to the instability of the Fe$_3$O$_4$ phase, some Fe is also present. Finally, an example of the X-ray Reflectivity (XRR) fits used to obtain film thickness is given in Fig. S2 (d).

**3.2 Characterization of the Fe$_3$O$_4$:Pt bilayers**
Additional characterization data for some of the Fe$_3$O$_4$:Pt bilayers is given in Fig. S3. Previous work showed that for $t_{Pt}$>2 nm a thin continuous paramagnetic Pt layer deposited on top of the Fe$_3$O$_4$ layer follows the wavy surface of the Fe$_3$O$_4$ layer. The Pt stacks on top of the Fe$_3$O$_4$ (111) planes and the two grains have an orientation relationship of [011]Pt//[011]Fe$_3$O$_4$, (111)Pt//(111)Fe$_3$O$_4$, which minimizes the interface energy due to minimal lattice mismatch of $d(111)_{Pt}$ (0.226 nm; JCPDS card 4-802) and $d(222)_{Fe3O4}$ (0.242 nm; JCPDS card 19-629).

XRD data for the sample series where the Fe$_3$O$_4$ and substrate thickness was varied is given in Fig. S3. For the substrate study, XRR (S3b) indicated similar Pt thickness (low frequency 'bumps' in the data) as well as some variation in the interface quality (whether high frequency fringes can be observed indicates the relative roughness of the substrate, $Fe_3O_4$ and Pt layers). Fig. S3 also shows the XRD and magnetization measurements of the sample series where the $Fe_3O_4$ thickness was varied. Of particular note is the saturation of the magnetic moment of $Fe_3O_4$ measured at 1 T for thicknesses > 80 nm.

### 3.3 Characterization of YIG

XRD of the YIG pellet presented here is given in Figure S4. No evidence of unreacted starter powders ($Fe_2O_3$, $Y_2O_3$) or potential secondary phase YIP ($YFeO_3$) was seen, as highlighted in Figure S4(b). Scanning electron microscopy images show grain structure with a porosity expected from 30% estimated open porosity as measured by the Archimedes method.

### 3.4 Additional spin Seebeck data

Fig. S5 shows some of the raw data for FP5 (analysed datapoints shown in Figure 2&3 of the main article). The heat source for these measurements was the same Peltier, with a maximum power output of approximately 1W. Fig. S6 shows some of the raw data for the YIG pellet as it was cut to different sizes. Fig. S7 shows more detailed data for the large area thin film measurement given in Figure 6 of the main text. Fig. S8 shows the results of the scribing study, where $A_T$ was fixed, but $L_x$, $L_y$ (not $L_x^T$, $L_y^T$) were varied by scribing away sections of the thin film, as illustrated in Fig. S8(a). In this case, the voltage always scaled with the contact separation, $L_y$, indicating that the behavior we see is not due to a change in the film properties (such as resistance), but scaling of the heat flux. Figure S9 shows additional spin Seebeck measurements where the rig was altered to monitor heat flux above and below the sample, with a resistor as the heat source. In this example, the Peltier area was 20x20 mm and the resistor was 57 Ohms. Measurements taken in air and under vacuum indicate a difference of approximately 3.9% due to convective losses (from the sides of the sample). Differences between $Q_{top}$ and $Q_{bottom}$ indicate radiative losses of the order of 1.3%, which is comparable to the estimate given in the main text. Finally, Figure S10 shows expanded data from Figure 6 of the main manuscript, where scatter in individual datapoints due to poor thermal contact, or variation in substrate thickness or area, which was not accounted for.

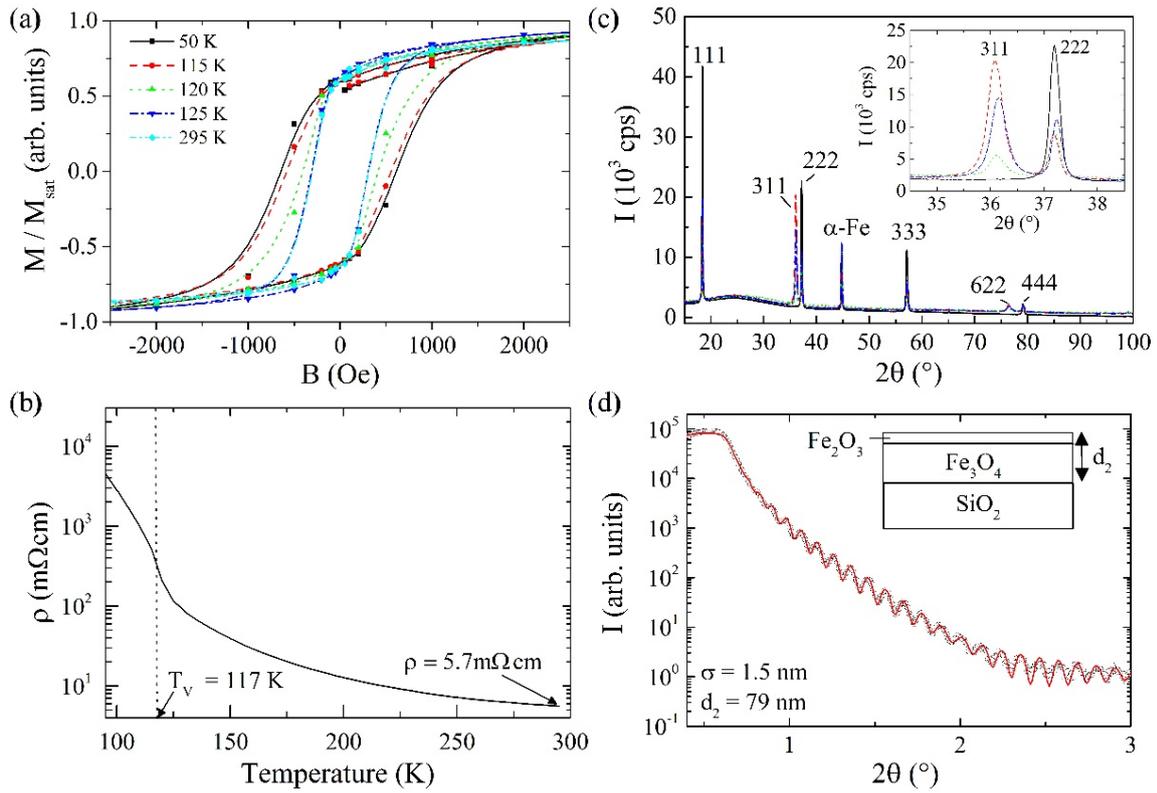

**FIG. S2** Characterization of the $Fe_3O_4$ film. a) SQUID magnetometry above and below the Verwey transition, $T_V$. b) Resistivity as a function of temperature. c) XRD of a set of 4 separately prepared $Fe_3O_4$ films. The inset shows a close-up of the (311), (222) peaks. d) Example XRR data (symbols) and fit (solid line), indicating thickness = 79 nm, roughness = 1.5 nm.

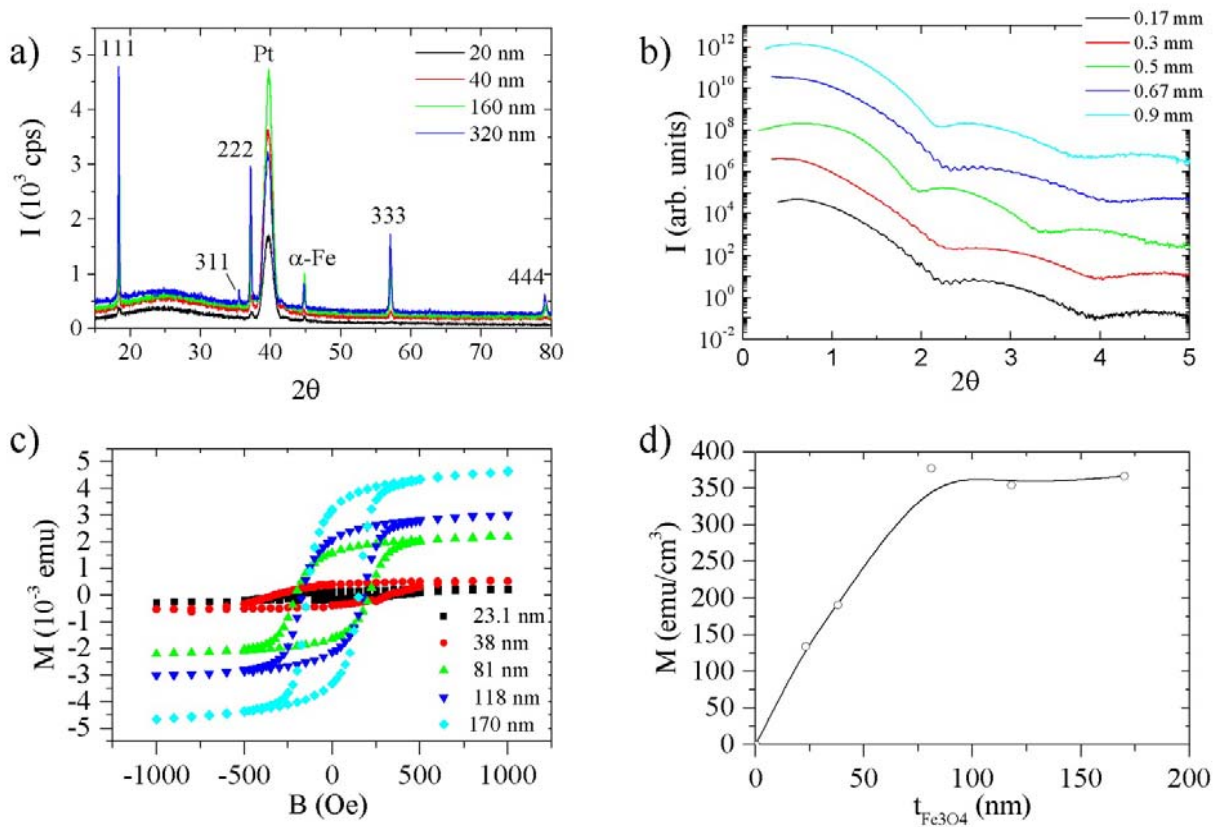

**FIG. S3** Characterization of the $Fe_3O_4$:Pt films, where $Fe_3O_4$ and substrate thickness was varied. a) XRD for $Fe_3O_4$ thickness study, b) XRR for substrate study, and c) SQUID magnetometry as a function of $Fe_3O_4$ thickness at 300K. d) Saturation magnetisation, M, as a function of $Fe_3O_4$ thickness.

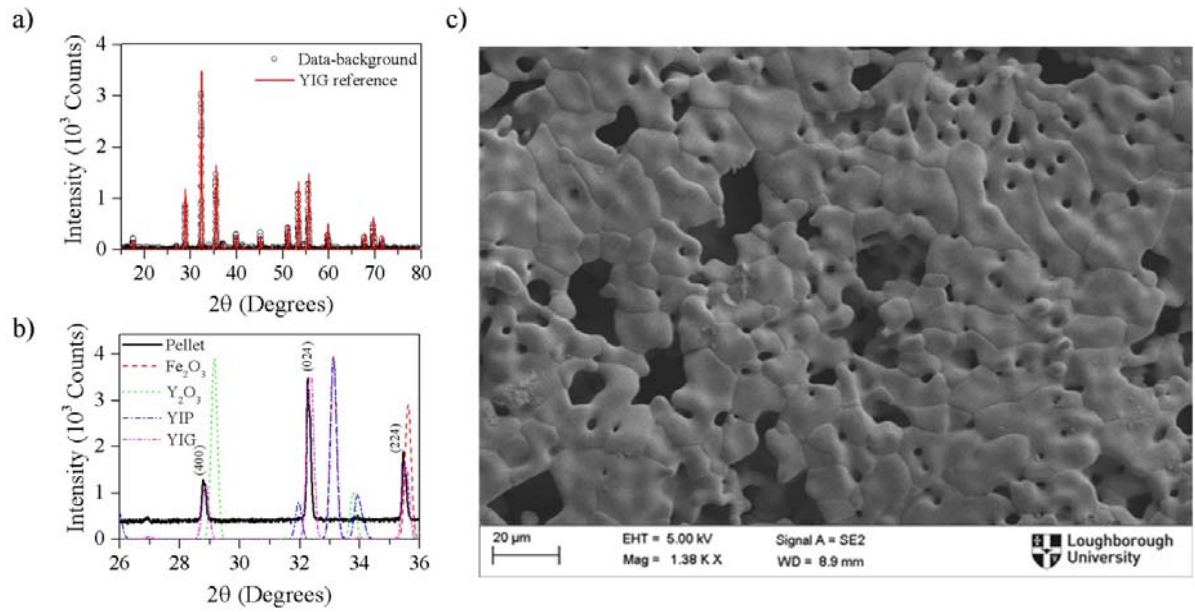

**FIG. S4** (a)&(b) X-ray diffraction data for the YIG pellet. (a) Normalized data alongside reference for $Y_3Fe_5O_{12}$ (YIG). (b) Zoomed in data selection plotted alongside possible impurity phases $Fe_3O_4$ and $Y_2O_3$ (starting powders) and $YFeO_3$ (YIP). (c) Scanning Electron Microscope image of the pellet, where grain size was found to be an average of 14.5 μm, with open porosity of 30%.

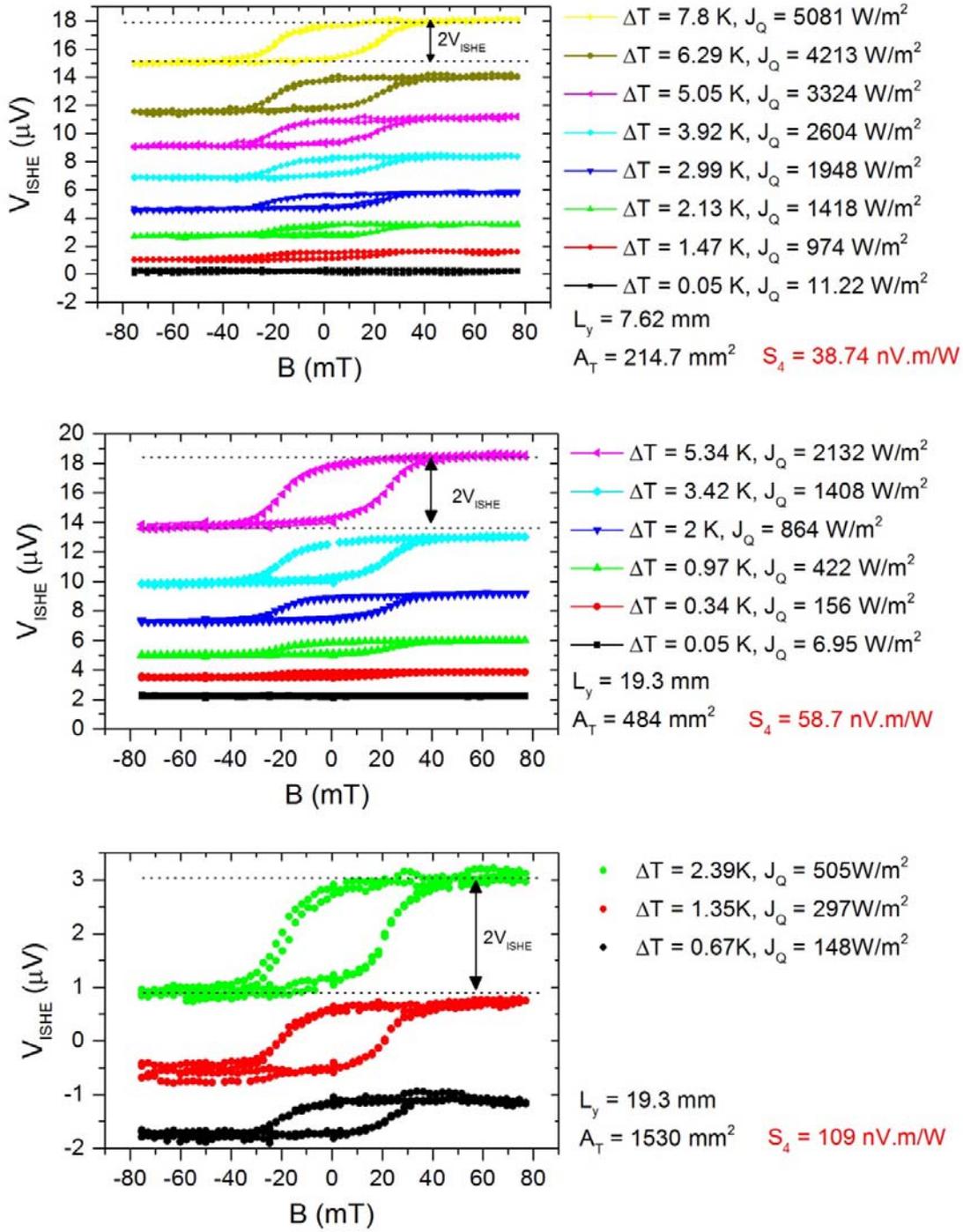

**FIG. S5** Raw voltage measurements for FP5 at 3 different areas, $A_T$. The contact separation, area, temperature difference, heat flux, $J_Q$, and resultant spin Seebeck coefficient, $S_4$, are shown for each sample. Data has been offset for clarity.

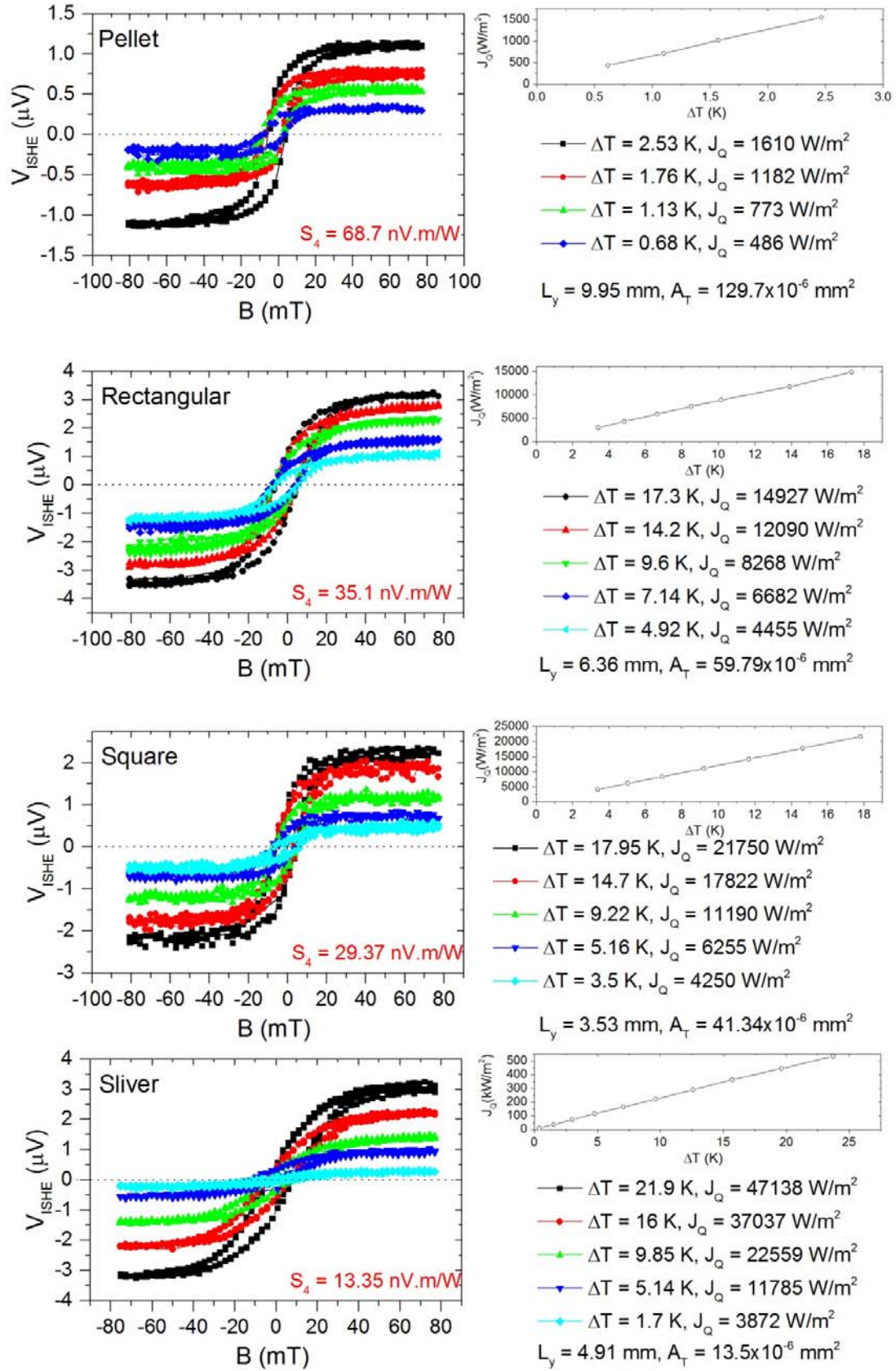

**FIG. S6** Raw voltage measurements for the bulk YIG pellet as it was cut down to different sizes, $A_T$. The contact separation, area, temperature difference, heat flux, $J_Q$, and resultant spin Seebeck coefficient, $S_4$, are shown for each sample. Inset shows the linear relationship between heat flux, $J_Q$, and temperature difference $\Delta T$, which starts to break down for $\Delta T > 10K$ due to radiation losses.

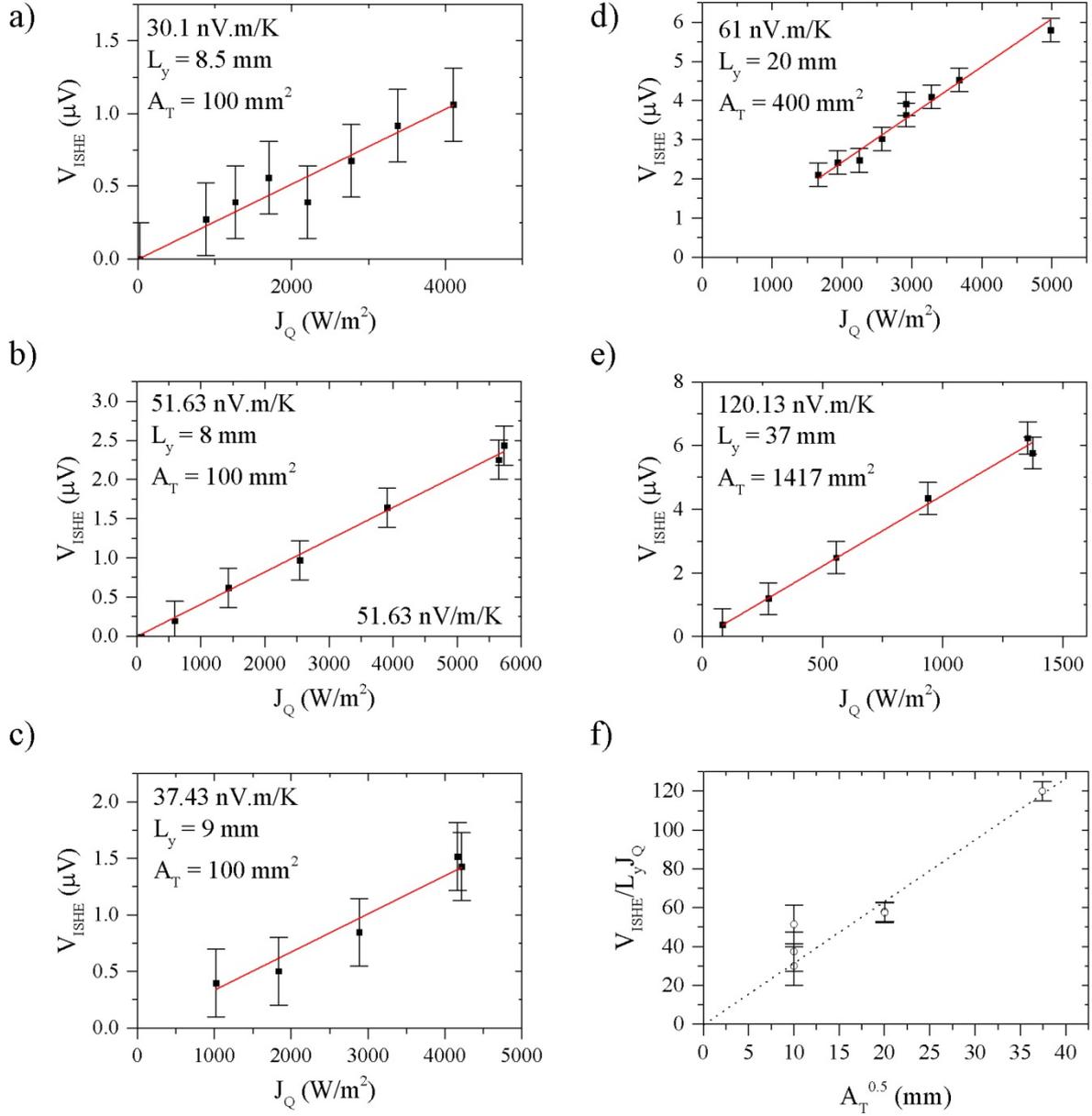

**FIG. S7** Summary of the voltage measured as a function of heat flux for the (a) – (c) 10x10, (d) 20x20, (e) 40x40 mm Peltier cell measurements. (f) S4 plotted as a function of $A_T^{0.5}$ for these samples. Note the scatter for the 100mm$^2$ samples is indicative of thickness variation over the 50x50 mm film, which was sampled by selecting pieces from 3 corners of the sample. Due to plume direction during the pulsed laser deposition process, we usually expect thickness variation of the Pt layer to differ by 10-15% at one edge of the film and this is evident by the larger voltage seen in (b), where the Pt thickness was approximately 0.5-1 nm thinner. To avoid this difference in standard measurements we constrain the sample deposition area to 40x40mm.

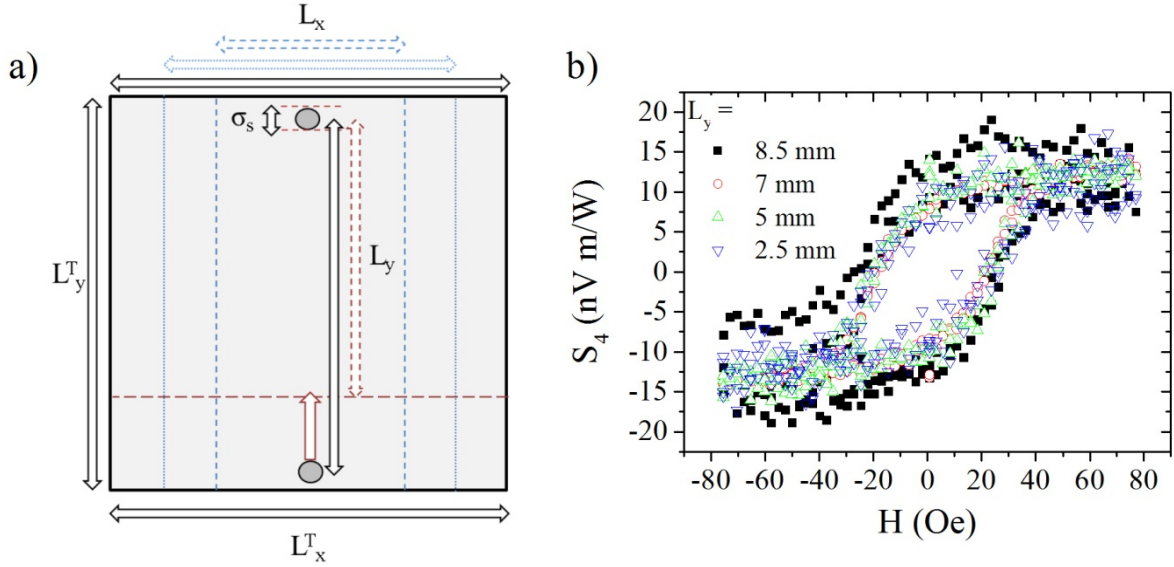

**FIG. S8** Results of the scribing study. As long as $A_T$ was unchanged, as $L_x$ and $L_y$ were varied (by scribing away the sample such that it is still thermally connected, but electrically isolated), the voltage always scaled the same. Large noise is due to reducing contact separation and additional electrical noise during some of the measurements (due to a loose wire).

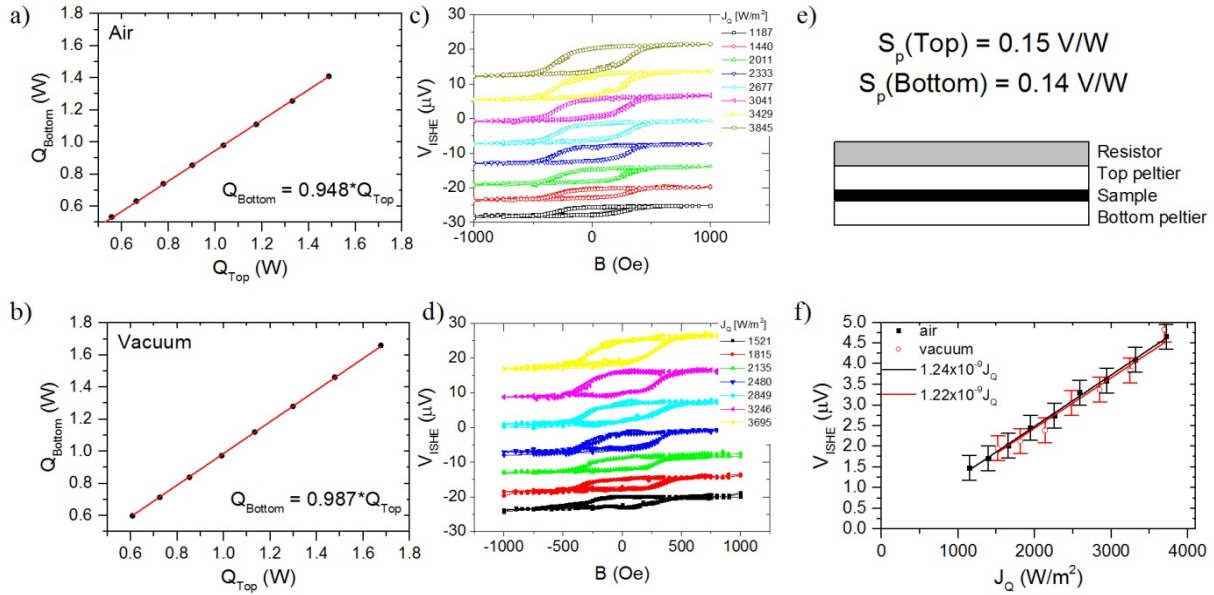

**FIG. S9** Additional spin Seebeck measurements for the 20x20 mm Peltier arrangement. (a) & (b) Final calibration data for the Peltier cells in air and under vacuum (P~1x10$^{-4}$ mBar), respectively. $Q_{top}$ is the heat measured by the top Peltier (W), and $Q_{bottom}$ is the heat measured by the bottom Peltier. This data indicates a heat loss of less than 6% across the stack. (c) & (d) Raw voltage measurements (data offset for clarity) at various heating powers (driven by a resistive heater) in air and vacuum, respectively. (e) schematic of the altered measurement. (f) $V_{ISHE}$ as a function of heat flux (extracted from data in (c) and (d)) for the air and vacuum measurements. The fit to this data indicates a deviation of approximately 2%.

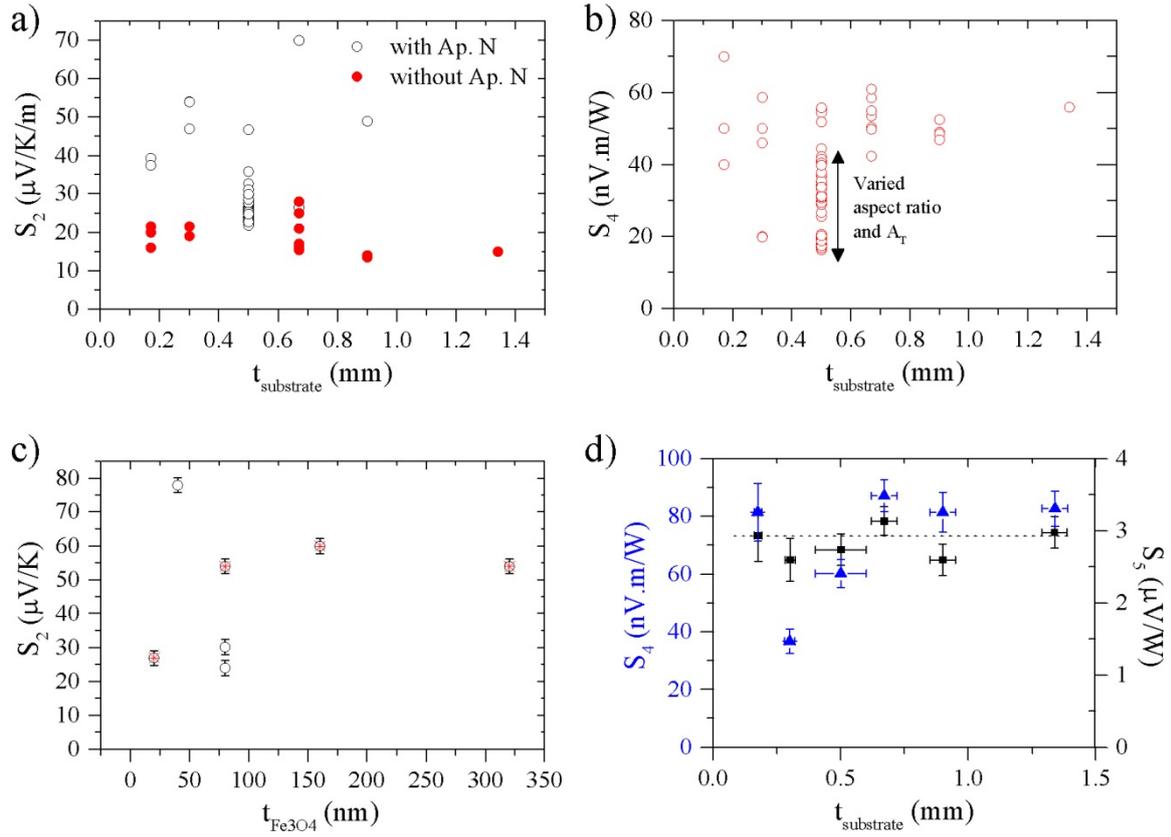

**FIG. S10** Data from the substrate and $Fe_3O_4$ thickness study plotted to show scatter in datapoints. (a) Impact of poor thermal contact on $S_2$, (b), Impact on $S_4$ of the variation of aspect ratio for the substrate measurements. (c) $S_2$ measured for the $Fe_3O_4$ thickness measurements. (d) $S_4$ (blue symbols) versus $S_5$ (black symbols) for the substrate measurements.

11